\documentclass[aps,twocolumn]{revtex4}
\usepackage{epsfig}
\usepackage{amsmath}
\def\bivec#1{\vbox{\ialign{##\crcr $\leftrightarrow$\crcr\noalign{
 \kern-1pt \nointerlineskip}$\hfil\displaystyle{#1}\hfil$\crcr}}}
\begin{document}

\title{Coupling Lattice Boltzmann with Atomistic Dynamics for the
multiscale simulation of nano-biological flows}

\author{Maria Fyta$^1$, Simone Melchionna$^2$, Efthimios Kaxiras$^1$,
and Sauro Succi$^3$}
\affiliation{$^1$ Department of Physics and School of Engineering and Applied
Sciences, Harvard University, Cambridge, MA, USA\\
$^2$ INFM-SOFT-CNR, Department of Physics, Universit\`a di Roma
\it{La Sapienza}, P.le A. Moro 2, 00185 Rome, Italy \\
$^3$ Istituto Applicazioni Calcolo, CNR,
Viale del Policlinico 137, 00161, Rome, Italy
}

\date{\today}

\begin{abstract}
We describe a recent multiscale approach based on the concurrent
coupling of constrained molecular dynamics for long biomolecules
with a mesoscopic lattice Boltzmann treatment of solvent
hydrodynamics. The multiscale approach is based on a simple scheme of 
exchange
of space-time information between the atomistic and mesoscopic
scales and is capable of describing self-consistent hydrodynamic
effects on molecular motion at a computational cost which scales
linearly with both solute size and solvent volume. For an
application of our multiscale method, we consider the much studied
problem of biopolymer translocation through nanopores: we find
that the method reproduces with remarkable accuracy the
statistical scaling behavior of the translocation process and
provides valuable insight into the cooperative aspects of biopolymer
and hydrodynamic motion.
\end{abstract}

\maketitle

\section{Introduction}

Modeling biological systems in an efficient and reliable way is a
delicate task, which calls for the development of innovative
computational methods, often requiring sophisticated upgrades and
extensions of techniques originally developed for physical and/or
chemical stand-alone applications. Indeed, biological systems
exhibit a degree of complexity and diversity straddling across
many decades in space-time resolution, to the point that, for many
years, biological systems served as a paradigm of the kind of
complexity which can only be handled in qualitative or descriptive
terms. Advances in computer technology, combined with constant
progress and breakthroughs in simulational methods, are closing
the gap between quantitative models and actual biological
behavior. The main computational challenge raised by biological
systems remains the wide and disparate range of spatio-temporal
scales involved in their dynamical evolution, with protein
folding, morphogenesis, intra- and extra-cellular communication,
being just a few examples.

In response to this challenge, various strategies have been
developed recently, which are in general referred to as {\it
multiscale modeling}. These methods are based on composite
computational schemes which rely upon multiple levels of
description of a given biological system, most typically the
atomistic and continuum levels. The multiple descriptions are then
glued together, through suitable 'hand-shaking' procedures, to
produce the final composite multiscale algorithm. To date, the
mainstream multiscale modeling is based on the coupling between
atomistic and continuum models. This choice reflects the
historical developments of statistical mechanics and computational
physics. In essence, continuum methods reduce the information to a
small number of distributed properties (fields), whose space-time
evolution is computed by solving a corresponding set of partial
differential equations, such as reaction-diffusion-advection
equations. Atomistic models, on the other hand, rely on the well
established approach of molecular dynamics, possibly including
extensions capable of dealing with a quantum description. It is
perhaps interesting to notice that this two-stage
continuum-to-atomistic representation overlooks a third,
intermediate, level of description, that is the mesoscopic level,
as typically represented by Boltzmann kinetic theory and its
extensions. Kinetic theory lies between the continuum and
atomistic descriptions, and it is thereby natural to expect that
it should provide an appropriate framework for the development of
robust multiscale methodologies.

Until recently, this approach has been hindered by the fact that
the central equation of kinetic theory, the Boltzmann equation,
was typically perceived computationally nearly as demanding as
molecular dynamics, and yet of very limited use for dense fluids
(water being the typical biological medium), due to the lack of
many-body correlations. Recent developments in lattice kinetic
theory \cite{LBE,LBEORIG} are making this view obsolete. Over the
last decade, such developments have provided solid evidence that
suitably discretized forms of minimal kinetic equations, and most
notably the Lattice Boltzmann equation, are giving rise to very
efficient algorithms capable of handling complex flowing systems
across many scales of motion. The behavior of fluid flow is
described through minimal forms of the Boltzmann equation, living
on a discrete lattice. The lattice dynamics is designed in such a
way as to reflect the basic conservation laws of continuum
mechanics, and also to host additional (mesoscopic) physics which
is not easily accommodated by continuum models. Remarkably, both
tasks can be achieved within the same algorithm, which proves
often computationally advantageous over the continuum approach
based on the Navier-Stokes equations.
% Moreover, lattice kinetic
% theory has proven capable of dealing with complex flows, such as
% flows with phase transitions and strong heterogeneities, for which
% continuum equations are either not established or just exceedingly
% difficult to solve (for a review see \cite{LGA05}). 
It is only
very recently that LB advances have started to be incorporated
within a new class of mesoscopic multiscale solvers \cite{MULBE}.
In this work, we address precisely such mesoscopic multiscale
solvers, with specific focus on the biologically important problem
of biopolymer translocation through nanopores.
Our procedure is based on the assumption that, in order to capture the
essential aspects of the translocation process, it is not necessary to
resolve all underlying atomistic details. As a result, both solvent
and solute degrees of freedom are treated through appropriate coarse-graining.
Due to the intrinsic coarse-graining nature of our methodology, the direct mapping to
experimental conditions has to proceed through the adjustment of appropriate 
parameters.

\section{Multiscale coupling methodology}

We will discuss the implementation of how a {\it mesoscopic} fluid
solver, the lattice Boltzmann method (LB), can be coupled
concurrently to the atomistic scale employing explicit atomistic
dynamics 
which, for simplicity, will be named molecular dynamics (MD) in 
a {\em broad sense}.  
This procedure involves different levels of the
statistical description of matter (continuum and atomistic) and is
able to handle different scales through the spatial and temporal
coupling between the constrained molecular dynamics for the
polymer evolution and the lattice Boltzmann treatment of the
explicit solvent dynamics. This multiscale framework is well
suited to address a class of biologically related problems.

The solvent dynamics does not require any form of statistical
ensemble averaging, as it is represented through a discrete set of
pre-averaged probability distribution functions (the
single-particle Boltzmann distributions), which are propagated
along straight particle trajectories. At variance with Brownian
dynamics, the lattice Boltzmann approach handles the
fluid-mediated solvent-solvent interactions through an explicit
representation of local collisions between the solvent and solute
molecules. By leveraging space-time locality, the corresponding algorithm
scales linearly with the number of beads, as opposed to the
(super)quadratic dependence of Brownian dynamics. This dual
field/particle nature greatly facilitates the coupling between the
mesoscopic and atomistic levels, both on conceptual and
computational grounds. Full details on this scheme are reported in
Ref. \cite{ourLBM}. It should be noted that, LB and MD
have been coupled before for the investigation
of single-polymer dynamics \cite{DUN}, but here this coupling is
extended to {\it long} molecules of biological interest.

A word of caution is in order with the Stokes limit $Re\rightarrow 0$.
In this limit, the scale separation between atomistic and hydrodynamic 
degrees of freedom becomes opaque, as the atomic mean free path
becomes comparable with hydrodynamic scales.
However, it has been shown that finite Reynolds corrections to hydrodynamics
have very negligible effects on the solvent-mediated forces between suspended bodies,
e.g. Oseen-level hydrodynamics is satisfactorily recovered 
\cite{LADDVERBERG,CATES,ourLBM,PREPA}.

We first turn to the atomistic part within our approach and
consider a polymer consisting of $N$ monomer units (also referred
to as beads). Each bead of the polymer is advanced in time
according to the following set of Molecular Dynamics
equations:
\begin{equation}
\vec{F}_{tot,i} = \vec{F}_{c,i} + \vec{F}_{drag,i}
+ \vec{F}_{r,i}+\vec{F}_{\kappa,i}
\label{eq:force}
\end{equation}
where the index $i$ runs over all beads. In this expression,
$\vec{F}_{c,i}$ is a conservative force describing bead-bead
interactions, represented here by a Lennard-Jones potential:
\begin{equation}
V_{LJ}(r) = 4 \varepsilon \left[ \left(\frac{\sigma}{r}\right)^{12}
- \left(\frac{\sigma}{r}\right)^{6}\right]
\end{equation}
This potential is truncated at a distance of r=$2^{1/6}\sigma$
and augmented by an angular harmonic term to account
for distortions of the angle between consecutive bonds:
 \begin{equation}
 V_{ang}(\phi) =\frac{\kappa_{\phi} \phi^2}{2}
 \label{Ang_potential}
 \end{equation}
with $\phi$ the relative angle between two consecutive bonds, and
$\kappa_{\phi}$ a constant. Torsional motions are not included in
the present model, but can easily be incorporated if needed. The
second term in Eq.(\ref{eq:force}), $\vec{F}_{drag,i}$, represents
the dissipative drag force due to polymer-fluid coupling given by
\begin{equation}
\vec{F}_{drag,i} =  -m \gamma (\vec{v}_i-\vec{u}_i)
\end{equation}
with $\vec{v}_{i}$, $\vec{u}_i$ the bead and fluid velocity
evaluated at the bead position $\vec{r}_i$ of bead $i$ with a mass
$m$; $\gamma$ is the friction coefficient. In addition to
mechanical drag, the polymer feels the effects of stochastic
fluctuations of the fluid environment, which is related to the
third term in Eq.(\ref{eq:force}), $\vec{F}_{r,i}$, an
uncorrelated random force with zero mean acting on bead $i$. The
term $\vec{F}_{\kappa,i}$ in Eq.(\ref{eq:force}) is the reaction
force resulting from $N-1$ holonomic constraints for molecules
modelled with rigid covalent bonds. The usage of constraints
instead of flexible bond lengths makes it possible to eliminate
unimportant high-frequency intra-molecular motion which might lead
to numerical instabilities. In order to avoid spurious
dissipation, the bead velocities are required to be strictly
orthogonal to the relative displacements. The constraints on both
positions and velocities are enforced over positions and momenta
separately via the SHAKE and RATTLE algorithms
\cite{SHAKE,RATTLE}.

\begin{figure*}
\begin{center}
\includegraphics[width=0.7\textwidth]{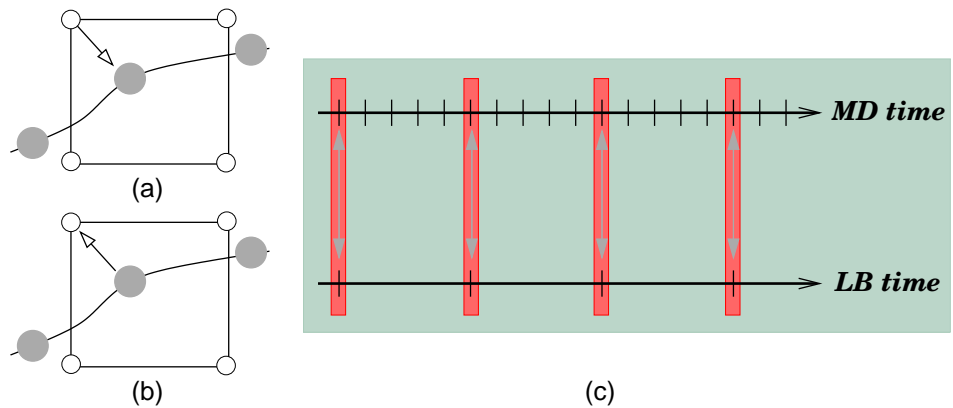}
\caption{\label{interpxch}Transfer of spatial information (a) from
grid to particle, and (b) from particle to grid. Grey spheres
denote beads, while in white are the lattice sites. In (c) the
information exchange (LB-MD couling) is sketched through the
vertical highlighted regions between the two different scales of
our multiscale approach. The MD marches in time on a finer
time-scale (by a factor of 5 in the sketch and our simulations)
than the LB solver.}
\end{center}
\end{figure*}

The LB equation is a minimal form of the Boltzmann kinetic
equation in which all details of molecular motion are removed
except those that are strictly needed to recover hydrodynamic
behavior at the macroscopic scale (mass-momentum and energy
conservation). The result is an elegant equation for the discrete
distribution function $f_p(\vec{x},t)$ describing the probability
to find a LB particle at lattice site $\vec{x}$ at time $t$ with a
discrete speed $\vec{c}_p$. Specifically, in this work we are
dealing with nanoscale flows and will consider the fluctuating
Lattice Boltzmann equation which takes the following form
\cite{adhikari}:
\begin{widetext}
\begin{equation}
f_p(\vec{x}+ \vec{c}_p \Delta t,t+\Delta t) = f_p(\vec{x},t) -
\omega \Delta t (f_p-f_p^{eq})(\vec{x},t) +  F_p \Delta t + S_p \Delta t
\label{lbe}
\end{equation}
\end{widetext}
The particles can only move along the links of a regular lattice
defined by the discrete speeds, so that the synchronous
displacements $\Delta \vec{x}_p = \vec{c}_p \Delta t$ never take
the fluid particles away from the lattice. For the present study,
the standard three-dimensional 19-speed lattice is used (see Fig.1
in Ref. \cite{MULBE}). The right hand side of Eq.~\ref{lbe}
represents the effect of intermolecular solvent-solvent
collisions, through a relaxation toward local equilibrium,
$f_p^{eq}$, typically a second order (low-Mach) expansion in the
fluid velocity of a local Maxwellian with speed $\vec{u}$:
\begin{equation}
f_p^{eq} = w_p \rho \lbrace
1 + \frac{\vec{u} \cdot \vec{c}_p}{c_s^2} +
\frac{1}{2c_s^4}  [\vec{u} \vec{u} :
(\vec{c}_p \vec{c}_p - c_s^2 {\rm \bivec{I}})]
\rbrace
\end{equation}
 where $c_s $ is the sound speed of the solvent,
$w_p$ is a set of
weights normalized to unity, {\rm \bf \bivec{I}} is the unit
tensor in configuration space, and $\rho$ is the local density.
The relaxation frequency $\omega$ controls the fluid kinematic
viscosity $\nu$, through the relation $\nu= c_s^2 (1/\omega-\Delta
t/2)$ \cite{MULBE}
and $\Delta t$ the LB time-step. Knowledge of the discrete
distributions $f_p$ allows the calculation of the local density
$\rho$, flow speed $\rho \vec{u}$ and momentum-flux tensor
$\bivec{P}$, by a direct summation upon all discrete
distributions:
\begin{eqnarray}
\rho(\vec{x},t)&=&\sum_{p} f_{p}(\vec{x},t)  \label{dens} \\
\rho \vec{u} (\vec{x},t)&=&\sum_{p} f_p(\vec{x},t) \vec{c}_p \label{vel} \\
\bivec{P} (\vec{x},t)&=&\sum_{p} f_p(\vec{x},t) \vec{c}_p \vec{c}_p
\label{pre}
\end{eqnarray}
The diagonal component of the momentum-flux tensor gives the fluid
pressure, while the off-diagonal terms give the shear-stress. Both
quantities are available locally and at any point in the
simulation. Thermal fluctuations are included through the source
term $F_p$ in Eq. (\ref{lbe}), which is consistent with the
fluctuation-dissipation theorem at {\em all} scales. In the same
equation, the polymer-fluid back reaction is described through the
source term $S_p$, which represents the momentum input per unit
time due to the reaction of the polymer on the fluid populations.
This back reaction is given by the following expression:
\begin{equation}
S_p (\vec{x},t)
= \frac{w_p}{c_s^2}
\sum_{i \in D(x)} [ \vec{F}_{drag,i} + \vec{F}_{r,i} ] \cdot \vec{c}_p
\end{equation}
where $D(x)$ denotes the mesh cell to which the {\it i}$^{th}$
bead belongs. All quantities in this equations have to reside on
the lattice nodes, thereby the frictional and random forces need
to be extrapolated from the particle to the grid location.

In the LB solver, free-streaming proceeds along straight
trajectories which secures exact conservation of mass and momentum
of the numerical scheme, but also greatly facilitates the
imposition of geometrically complex boundary conditions. There is
no need to solve the computationally expensive Poisson equations,
since the pressure field is locally available. All interactions
are local, rendering the LB scheme ideal for parallel computing.
More advanced Lattice Boltzmann models \cite{karlin} also have
been developed and could equally well be suited for coupling to
atomic scale dynamics.

The Molecular Dynamics solver is marched in time with a
stochastic integrator (due to extra non-conservative and random
terms) \cite{melchJCP}, proceeding at a fraction $1/M$ of the LB time-step $\Delta
t$: $\Delta t_{MD} = \Delta t /M$. The time-step ratio $M>1$
controls the scale separation between the solvent and solute
timescales and should be chosen as small as possible, consistent
with the requirement of providing a realistic description of the
polymer dynamics. The MD cycle is repeated $M$ times, with the
hydrodynamic field frozen at each LB time-stamp $t_n= n \Delta t$,
at which the transfer of spatial information from grid to particle
locations (and conversely) is performed. For this transfer, a
simple nearest grid point interpolation scheme is used (see
Fig.\ref{interpxch}(a), (b)), on account of its simplicity. At a
time step $t_n=n \Delta t$, the pseudo-algorithm describing single
LB time-step, will be:
\begin{enumerate}
\item Interpolation of the velocity: $\vec{u}(\vec{x}) \rightarrow \vec{u}_i$
\item For $k=1,M$:
%\begin{itemize}
%\item[]
Advance the molecular state from $t$ to $t+\Delta t_{MD}$
%\end{itemize}
\item Extrapolation of the forces: $\vec{F_i} \rightarrow \vec{F}(\vec{x})$
\item Advance the Boltzmann populations from $t$ to $t+\Delta t$
\end{enumerate}
A sketch of this scheme is presented in Fig.~\ref{interpxch}. In
terms of computational efficiency, $\Delta t_{MD}$ is largely
independent of the number of beads $N$ because (i) the LB-MD
coupling is local, (ii) the forces are short ranged and (iii) the
SHAKE/RATTLE algorithms are empirically known to scale linearly
with the number of constraints. Up to this point we described a
scheme that is general and applicable to any situation where a
long polymer is moving in a solvent. This motion is of great
interest for a fundamental understanding of polymer dynamics in
the presence of the solvent and crucial in relevant biophysical
processes.

\section{Biopolymer translocation through nanopores}

We next turn to an application of the multiscale scheme described
in the previous section. We were motivated by recent experimental
studies which have focused on the translocation of biopolymers
such as RNA or DNA through nanometer sized pores. These explore
{\it in vitro} the translocation process through micro-fabricated
channels under the effects of an external electric field, or
through protein channels across cellular membranes \cite{EXPRM}.
In particular, recent experimental work has focussed on the
possibility of fast DNA-sequencing through electronic means,  that
is, by reading the DNA sequence while it is moving through a
nanopore under the effect of a localized electric
field \cite{EXPRM}. This type of biophysical processes are
important in phenomena like viral infection by phages,
inter-bacterial DNA transduction or gene therapy \cite{TRANSL}.
Some universal features of DNA translocation have also been
analyzed theoretically, by means of suitably simplified
statistical schemes \cite{statisTrans}, and non-hydrodynamic
coarse-grained or microscopic models \cite{DynamPRL,coarse}.
However, these complex phenomena involve the competition between
many-body interactions at the atomic or molecular scale,
fluid-atom hydrodynamic coupling, as well as the interaction of
the biopolymer with wall molecules in the region of the pore.
Resolving these interactions is essential in understanding the
physics underlying the translocation process. To this end, we
model the dynamics of biopolymer translocation through narrow
pores, using the multiscale scheme described above.

\begin{figure*}
\begin{center}
\includegraphics[width=0.6\textwidth]{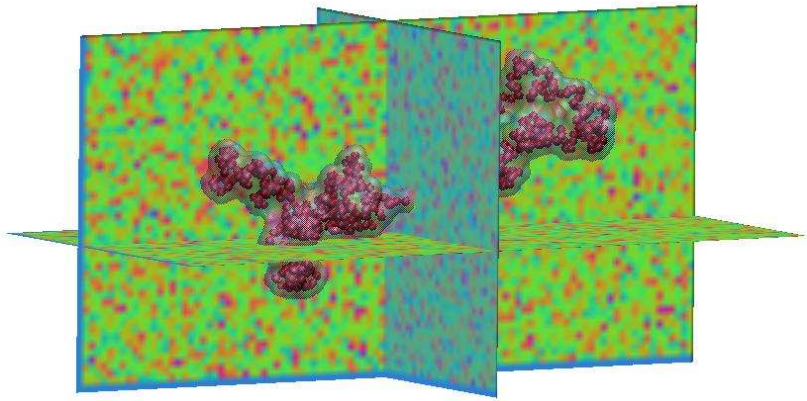}
\caption{\label{3dbox} A typical translocation event: The 3-D box
including the fluid and the polymer, as well as the wall (not shown)
that separates the box into two chambers. A polymer (red beads)
representing double-stranded DNA is shown at an instant where
about $60$\% of the beads have translocated. The fluid is also
shown as a 3-D representation of the velocity magnitude at the
vicinity of the beads and on 2-D planes (red denotes high value).
Translocation is induced by means of a pulling force $F_{drive}$
with a direction from right to left.}
\end{center}
\end{figure*}

Our numerical simulations are performed in a three-dimensional box
of size $N_x \times N_y  \times N_z $ in units of the lattice
spacing $\Delta x$. The box contains both the solvent and the
polymer. We take $N_x = 2 N_y$, $N_y = N_z$; a separating wall is
located in the mid-section of the $x$ direction, at $x= N_x/2$.
For polymers with less than $500$ beads we use $N_x = 80$ while
for larger polymers $N_x = 100$. At $t=0$ the polymer resides
entirely in the right chamber at $x> N_x/2$. At the center of the
separating wall, a square hole of side $d=3\Delta x$ is opened up,
through which the polymer can translocate from one chamber to the
other. A 3-D representation of a typical translocation event is
shown in Fig.\ref{3dbox}. The magnitude of the fluid speed is also
mapped on different planes and as a 3-D contour surface
surrounding the polymer beads.

We elaborate next on the main parameters involved in the
simulation (additional details are provided in Ref.
\cite{ourLBM}). All parameters are measured in units of the
lattice Boltzmann time-step $ \Delta t$ and spacing $\Delta x$,
which are set equal to 1. The parameters for the Lennard-Jones
potential are $\sigma=1.8 $, and $\epsilon=2 \times 10^{-3}$ and
the bond length among the beads is set to $b=1.2 $. The solvent
density and kinematic viscosity are $1$ and $0.1$, respectively,
and the inverse temperature is $\beta=10^{4}$. It must be noted
that the friction coefficient taken as $\gamma=0.1$ is a parameter
governing both the structural relation of the polymer towards
equilibrium and the strength of the coupling with the surrounding
fluid. The MD time-step is a $1/M$ fraction of the LB time-step,
as mentioned previously, and we set $M=5$.

Translocation is induced by a constant electric force,
$\vec{F}_{drive,i}$, which acts along the $x$ direction and is
confined in a rectangular channel of size $3\Delta x \times \Delta
x \times \Delta x$ along the streamwise ($x$ direction) and
cross-flow ($y,z$ directions). This force is included as an
additional term in Eq.(\ref{eq:force}), and is the driving force
representing the effect of the external field in the experiments.
We use $F_{drive,i} =0.02$. This choice of parameter values
implies that we are describing the {\it fast} translocation
regime, in which the translocation time is much smaller than the
Zimm time, which is the typical relaxation time of the polymer
towards its native (minimum energy, maximum entropy)
configuration. Under these conditions, the many-body aspects of
the polymer dynamics cannot be ignored because the beads along the
chain do not move independently.

Here, we model DNA as a polymeric chain of a number of segments
(the beads) and trace its dynamic evolution interacting with a
fluid solvent as it passes through a narrow hole that is
comparable with the bead size. Each bead maps to a number of
base-pairs (bp), ranging from about 8 (similar to the hydrated
diameter of B-DNA in physiological conditions) to $\sim10^3$
\cite{beadParam1,beadParam2}. In order to estimate this number for
our simulations and interpret our results in terms of physical
units we examine the persistence length ($l_p$) of the
semiflexible polymers used in our simulations. We use the formula
for the fixed-bond-angle model of a worm-like chain \cite{lpWCL}:
\begin{equation}
l_p=\frac{b}{1-\cos\langle \theta \rangle}
\end{equation}
where $\langle\theta\rangle$ is complementary to the average bond
angle between adjacent bonds. In lattice units ($\Delta x$) an
average persistence length for the polymers considered was found
to be approximately $12$. For $\lambda$-phage DNA, $l_p\sim 50$ nm
\cite{lpDNA}, which is set equal to $l_p$ for our polymers.
Thereby, the lattice spacing is $\Delta x \sim 4$ nm, which is
also the size of one bead. Given that the base-pair spacing is
$\sim0.34$ nm, one bead maps to approximately $12$ base pairs.
With this mapping, the pore size is about $\sim 12$ nm, close to
the experimental pores which are of the order of $10$ nm. The
polymers presented here with $N=20-500$ beads correspond to DNA
lengths in the range $0.2-6 \times 10^3$ bp. The DNA lengths used
in the experiments are larger (up to $\sim 10^5$ bp); with
appropriate computational resources, our multiscale scheme could
handle these lengths.

\begin{figure}
\begin{center}
\includegraphics[width=0.3\textwidth]{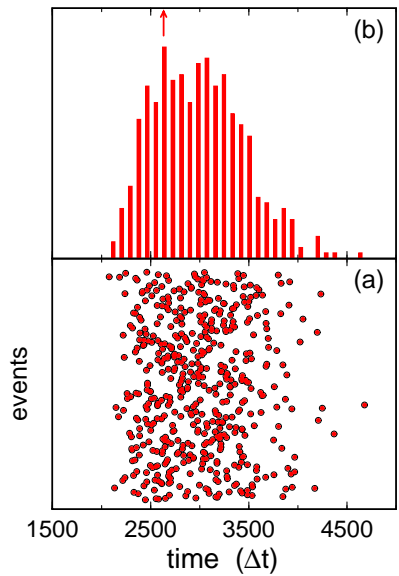}
\caption{\label{events} (a) The variety of different translocation
events and the corresponding translocation times; (b) from the
events in (a) the duration histogram is extracted for a molecule
with 100 beads, which represents double-stranded DNA with $1.2$
kbp. Time is given in units of the LB time-step $\Delta t$.}
\end{center}
\end{figure}

\begin{figure*}
\begin{center}
\epsfig{file=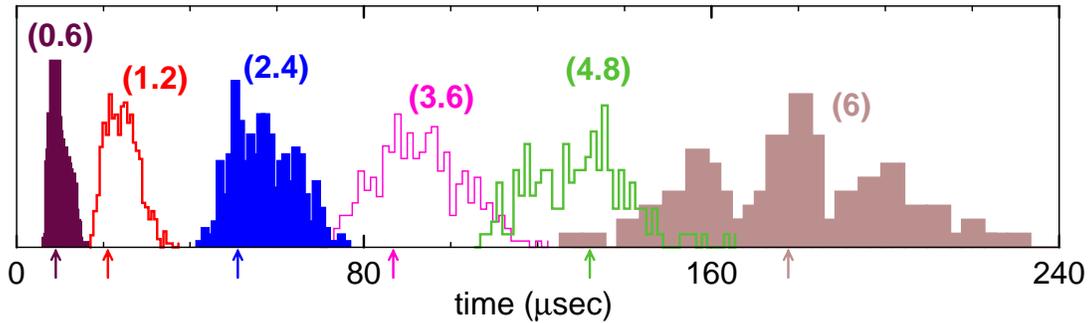,width=0.8\textwidth}
\caption{\label{histos}Histograms of calculated translocation
times for a large number of events and various DNA lengths shown
in parentheses in kbp ($10^3$ bp). The arrows point to the most
probable time for each polymer size.}
\end{center}
\end{figure*}

Having established the quantitative mapping of DNA base-pairs to
the simulated beads, we seek a comparison of the statistical
features of the simulated translocation process to experimental
studies. The ensemble of our simulations is generated by different
realizations of the initial polymer configuration to account for
the statistical nature of the process. We performed extensive
simulations of a large number of translocation events over
$100-1000$ initial polymer configurations for each polymer length.
The various events for a given length are depicted in Fig.
\ref{events}(a). The projected duration histograms are shown in
Fig. \ref{events}(b) in LB units. Similar distributions were
obtained for all the polymer lengths considered here, by
accumulating all events for each length. By choosing lengths that
match experimental data we compare the corresponding experimental
duration histograms (see Fig.~1c of Ref.~\cite{NANO}) to the
theoretical ones. This comparison sets the LB time-step to $\Delta
t\sim 8$ nsec. In Fig.~\ref{histos} we show the time distributions
for representative DNA lengths simulated here. In this figure,
physical units are used according to the mapping described above
for direct comparison to similar experimental data \cite{NANO}.
The MD time-step for $M=5$ will then be $\Delta t_{MD}\sim 1.6$
nsec indicating that the MD timescale related to the
coarse-grained model that handles the DNA molecules is
significantly stretched over the physical process. Exact match to
all the experimental parameters is of course not feasible with
coarse-grained simulations. Nevertheless, essential features of
DNA translocation are reasonably well reproduced, allowing the use
of the current approach to model similar biophysical processes
that involve biopolymers in solution. This can become more
efficient by exploiting the freedom of further fine-tuning the
parameters used in the multiscale model.

\begin{figure}
\begin{center}
\includegraphics[width=0.45\textwidth]{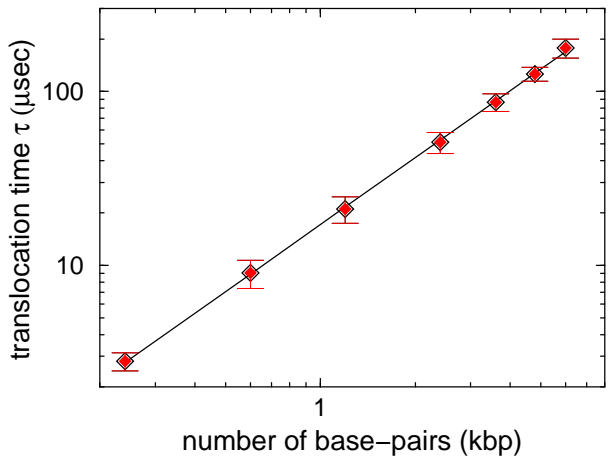}
\caption{\label{scaling} Scaling law of the translocation time
with the DNA length.}
\end{center}
\end{figure}

The variety of all the different initial polymer realizations
produce a scaling law dependence of the translocation times on
length \cite{coarse,NANO}. The duration histograms are not simple
gaussians, but are rather skewed towards longer times.
Accordingly, we use the most probable time (peak of the
distribution shown by the arrow in Fig.~\ref{events}(b)) as the
representative translocation time for every distribution; this is
also the definition of the translocation time in experiments, to
which we compare our results. Calculating the most probable times
for each length leads to the nonlinear relation between the
translocation time $\tau$ and the number of beads $N$: $\tau (N)
\propto N^{\alpha}$, with an exponent $\alpha \sim 1.28\pm 0.01$.
The scaling law is reported in Fig.~\ref{scaling} and is in very
good agreement with a recent experimental study of double-stranded
DNA translocation, that reported $\alpha\simeq1.27\pm0.03$
\cite{NANO}. In the absence of a solvent the exponent rises to
$1.36\pm0.03$. Such a difference indicates a significant
acceleration of the process due to hydrodynamic interactions.

\subsection{Dynamics of the translocation process}

We next turn to the dynamics of the biopolymer as it passes
through the pore. The simulations confirm that the polymer moves
through the pore in the form of two almost compact blobs on either
side of the wall. One of the blobs (the untranslocated part,
denoted by $U$) is contracting and the other (the translocated
part, denoted by $T$) is expanding. This behavior is visible in
Fig.~\ref{3dbox} for a random event and holds throughout the
process, apart from the end points (initiation and completion of
the translocation). A radius of gyration $R_I(t)$ (with $I=U,T$)
can be assigned to each of these blobs, following a static scaling
law with the number of beads $ N_{I}$: $R_I(t) \sim
N_{I}^{\mu}(t)$ with $\mu \simeq 0.6$ being the Flory exponent for
a three-dimensional self-avoiding random walk. Based on the
conservation of polymer length, $N_U+N_T = N$, an effective
translocation radius can be defined as
\begin{equation}
R_{E}(t)=[R_T^{1/\mu} (t)+R_U^{1/\mu}(t)]^{\mu}
\end{equation}
which should be constant when the static scaling applies. We
deduce from our simulations that $R_E(t)$ is approximately
constant for all times throughout the process except near the end
points, at which the polymer can no longer be represented as two
uncorrelated compact blobs and the static scaling no longer holds.

\begin{figure}
\begin{center}
\includegraphics[width=0.4\textwidth]{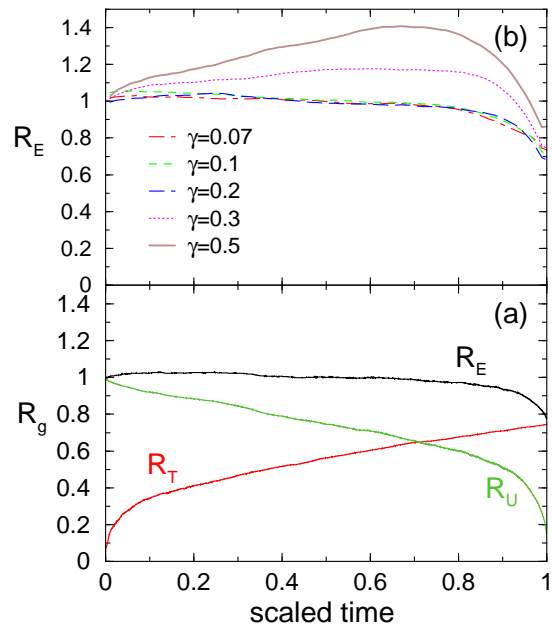}
\caption{\label{rgyr}
(a) Radii of gyration (translocated, untranslocated
and effective) for a polymer representing $4.8$ kbp as a function of time.
(b) Effective radii of gyration with time for different values of the
coupling $\gamma$ ($N=200\sim 2.4~kbp$). Time and $R_E$ are scaled with
respect to the total translocation time and $R_U(t=0)$ for each case.}
\end{center}
\end{figure}

In Fig.~\ref{rgyr}(a), we represent the time evolution of all
radii as averages over hundreds of events for a specific polymer
length. The time shown is scaled so that $t=1$ denotes the total
translocation time for an event. By definition, $R_U(t)$ vanishes
at $t=1$, while $R_T$ increases monotonically from $t=0$ up to
$t=1$, although it never reaches the value $R_U(t=0)$ (we
elaborate on this below).

It is essential to check the validity of the static scaling with
respect to the strength of the hydrodynamic field. To this end,
the effective radii of gyration are further explored in relation
to the parameter $\gamma$. We fixed the length to $N=200$ beads
and generated about 100 different initial configurations for each
value of $\gamma$. We present the variation of $R_E$ with this
coefficient in Fig. \ref{rgyr}(b): It is clearly visible in this
figure that $R_E$ is almost constant for small $\gamma$ but as
$\gamma$ increases, the radii are no more constant not only at the
end points, but also throughout the translocation. Large values of
the parameter $\gamma$ are interpreted as a strong molecule-fluid
coupling. The influence of the fluid on the beads, experienced by
the back reaction, is large and suppresses the polymer
fluctuations in such a way that the translocating biopolymer can
no longer be represented as a pair of compact blobs, and the
static scaling no longer holds.

Inspection of all the biopolymers at the end of the event reveals
that they become more compact after their passage through the
pore. This is quantitatively checked through the values of the
radii of gyration: the radius of gyration is considerably smaller
at the end than it was initially: $R_T(t=1)<R_U(t=0)$. The fact
that as the polymer passes through the pore it becomes more
compact than it was at the initial stage of the event may be
related to incomplete relaxation, but this remains to be
investigated. In Fig. \ref{rgyr}(b), all effective radii of
gyration at the final stage of the translocation decrease to a
value smaller than the initial $R_U(t=0)$. The ratio
$R_T(t=1)/R_U(t=0)$ is always smaller than 1 and ranges from
$0.72$ for $\gamma$=0.1 to $0.90$ for $\gamma$=0.5.

Throughout its motion the polymer continuously interacts with the
fluid environment. The forces that essentially control the process
are the electric drive $\vec{F}_{drive,i}$ and the hydrodynamic
drag $\vec{F}_{drag,i}$ which act on each bead. However, at the
end points (initiation and completion of the passage through the
pore) entropic forces become important \cite{blob}. The
fluctuations experienced by the polymer due to the presence of the
fluid are correlated to these entropic forces which, at least close to equilibrium, can be
expressed as the gradient of the free energy with respect to the
fraction of translocated beads. At the final stage of a
translocation event, the radius of the untranslocated part
undergoes a visible deceleration (see Fig.~\ref{rgyr}(a)), the
majority of the beads having already translocated. It is, thus,
entropically more favorable to complete the passage through the
hole rather than reverting it, that is, the entropic forces
cooperate with the electric field and the translocation is
accelerated.

The entropic forces can also lead to rare events, such as
retraction, which occur in our simulations at a rate less than 2\%
and depend on length, initial polymer configuration and parameter
set. A retraction event is related to a polymer that
anti-translocates after having partially passed through the pore.
We have visually inspected the retraction events and associate
them with the translocated part entering a low-entropy
configuration (hairpin-like) subject to a strong entropic
pull-back force from the untranslocated part: The translocated
part of the polymer assumes an elongated conformation, which leads
to an increase of the entropic force from the coiled,
untranslocated part of the chain. As a result, the translocation
is delayed and eventually the polymer is retracted.

\begin{figure}
\begin{center}
\includegraphics[width=0.4\textwidth]{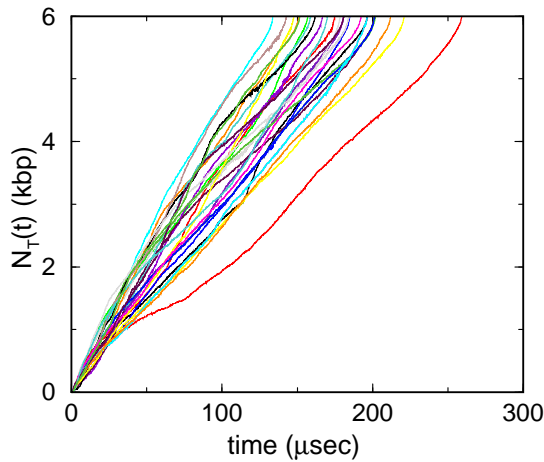}
\caption{\label{ntransl} Time evolution of the number of
translocated beads $N_T (t)$ for DNA with $6$ kbp. Curves
correspond to translocation events for a variety of different
initial configurations.}
\end{center}
\end{figure}

The fact that the entropic forces are related to the number of
translocated monomers $N_T(t)$ led us to investigate in more
detail the time evolution of this quantity. The number of
translocated monomers is plotted in Fig.\ref{ntransl} for various
initial configurations of the polymer. Each curve corresponds to a
different completed translocation event. The translocation for a
given polymer proceeds along a curve closely related to its
initial configuration and its interactions with the fluid; each
polymer follows a distinct trajectory as indicated by the variety
of curves. It is not possible to predict the ensuing behavior from
the initial polymer configuration in a simple and unique manner.

\subsection{Energetics of the translocation process}

\begin{figure}
\begin{center}
\includegraphics[width=0.45\textwidth]{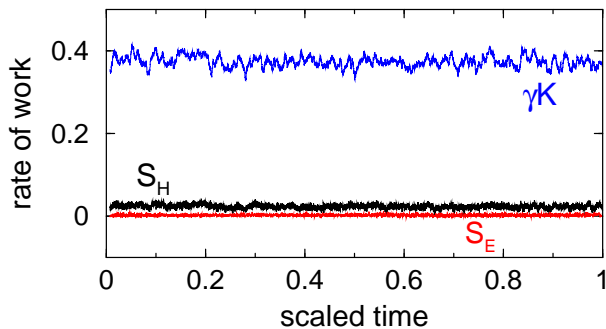}
\caption{\label{work}Rates of work ($S_{H}$, $S_{E}$) and kinetic
energy ($\gamma K$) as functions of time for one translocation
event of a polymer consisting of $N=300$ beads ($3.6$ kbp). Time
is scaled with respect to the total time of this event.}
\end{center}
\end{figure}

As a final step, we study the work performed on the biopolymer
throughout its translocation. On general grounds, hydrodynamic
interactions are expected to minimize frictional effects and form
a cooperative background that assists the passage of the polymer
through the pore. We investigate the cooperativity of the
hydrodynamic field through the {\it synergy} factor $S_H(t)$
defined as the work made by the fluid on the polymer per unit
time:
\begin{equation}
S_{H}(t)=\frac{dW_{H}}{dt}=\gamma \Big\langle\sum_i^{N}
\vec{v}_i(t) \cdot \vec{u}_i(t)\Big\rangle \label{eq:Shydro}
\end{equation}
where $W_{H}$ is the work of the fluid on the polymer. Through this
definition, positive values of this hydrodynamic work rate
indicate a cooperative effect of the solvent, while negative
values indicate a competitive effect by the solvent. The work done
per timestep by the electric field ($S_E(t)$) on the polymer can
also be easily obtained through the expression:
\begin{equation}
 S_E (t)=\frac{dW_E}{dt} =
 \Big\langle\sum_i^{N} \vec{F}_{drive,i} \cdot \vec{v}_{i}(t)\Big\rangle
\label{eq:Selec}
\end{equation}
where $W_{E}$ is the work of the electric drive on the polymer.
The brackets in Eq.(\ref{eq:Shydro}) and (\ref{eq:Selec}) denote
averages over different realizations of the polymer for the same
length. The results for the averages over all realizations are
qualitatively similar to the work rates for an individual event of
the same length. For all lengths studied here, we found that the
total work per timestep of the hydrodynamic field ($S_H(t)$) on
the whole chain is essentially constant, as shown in Fig.~\ref{work} 
for an individual event. For the same event, $S_E(t)$
is also constant with time. In the same figure the kinetic energy
$K$ of the polymer (plotted as $\gamma K$) for the same event is
shown for comparison. The kinetic energy is also constant with
time, as expected since the temperature in the simulations is held
constant, but its fluctuations differ from those of $S_E$ and
$S_H$. The larger value of $K$ with respect to both $S_H,~S_E$ can
be justified by the fact that the bead velocities are larger
compared to the fluid velocity. The hydrodynamic work per time is
also larger than the corresponding electric field work, because
the latter only acts in the small region around the pore. The
average of all these quantities over all events for the same
polymer length are also constant, and show smaller fluctuations
with time than those of any individual event.

\begin{figure}
\begin{center}
\includegraphics[width=0.4\textwidth]{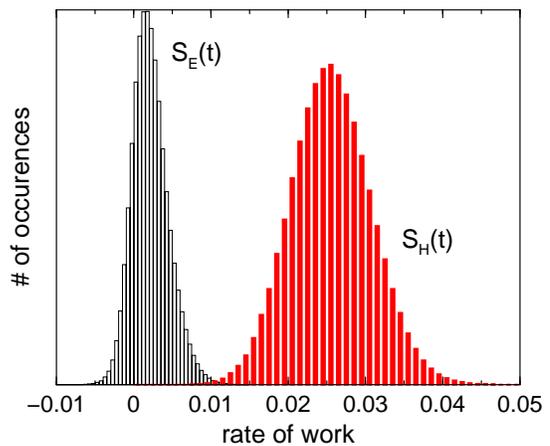}
\caption{\label{vargamma} Distribution of the translocation work
per unit time averaged over all events for a molecule with $3.6$
kbp. The contributions from both the hydrodynamic and electric
field are shown: $S_H(t)$ and $S_E(t)$, respectively.}
\end{center}
\end{figure}

In addition to the variation of the work rates with time it is
useful to analyze their distributions during translocation events.
We show these in Fig. \ref{vargamma}, where it is evident that the
distribution of $S_{H}(t)$ lies entirely in the positive range,
indicating that hydrodynamics turns the solvent into a cooperative
environment (it enhances the speed of the translocation process).
In the same figure, the distribution for $S_E(t)$ over all events
for the same length is also shown; this distribution is mostly
positive but has a small negative tail which indicates that beads
can be found moving against the electric field.

\subsection{Performance data}

We turn next to some technical aspects of or multiscale
simulations. The total cost of the computation scales roughly like
$$t \sim (t_{LB} V  + t_{MD} M N) N_{LB}$$
where $t_{LB}$ is the CPU time required to update a single LB site
per timestep and $t_{MD}$ is the CPU time to update a single bead
per timestep (including the overhead of LB-MD coupling); $V$ is
the volume of the computational domain in lattice units and $N$ is
the number of polymer beads, with $M$ the LB-MD time-step ratio;
$N_{LB}$ is the number of LB timesteps. Due to the fact that the
LB-MD coupling is local, the forces are short ranged and the
SHAKE/RATTLE algorithms are empirically known to scale linearly
with the number of constraints, so that $t_{MD}$ is largely
independent of $N$. The LB part is known to scale linearly with
the volume occupied by the solvent. Indeed, at {\em constant
volume}, the CPU cost of the simulations scales linearly with the
number of beads: The execution times for $50$, $100$ and $400$
beads are $0.433$, $0.489$, and $0.882$ sec/step, respectively on
a 2GHz AMD Opteron processor. By excluding hydrodynamics, these
numbers become $0.039$, $0.075$, and $0.318$ sec/step. For the
case where polymer {\em concentration} is kept constant, the
volume needed to accommodate a polymer of $N$ beads should scale
approximately as $N^{1.8}$. A typical translocation event with
$500$ beads, evolves over $30,000$ LB steps or $150,000$ MD steps.
Assuming $250$ flops/site/LB-step and $2500$ flops/bead/MD-step,
the previous equation leads to a computational cost of $\sim 6$
hrs. This is comparable with the time observed directly from the
simulations ($\sim 7$ hrs).

\section{Conclusions}

We have presented a new multiscale methodology based on the direct
coupling between atomistic motion and mesoscopic hydrodynamics of
the surrounding solvent. Due to the particle-like nature of the
mesoscopic lattice Boltzmann solver, this coupling proceeds
via simple interpolation/extrapolation in space and subcycling over time.
Correlations between the atomistic
and hydrodynamic scales are also explicitly included through
direct and local interactions between the solvent meso-molecules
and the polymer molecules. As a result, hydrodynamic interactions
between the polymer and the surrounding fluid are explicitly taken
into account, with no need of resorting to non-local
representations, such as the long-range Oseen tensor used in
Brownian dynamics. This allows a state-of-the-art modeling of
biophysical phenomena, where hydrodynamic correlations play a
significant role.

We have successfully applied our multiscale methodology to the
problem of biopolymer translocation through nanoscale pores.
Besides statistical properties, such as scaling exponents, the
present methodology affords direct insights into the details of
the {\it dynamics} as well as the {\it energetics} of the
translocation process, thereby offering a very valuable complement
to experimental investigations of these complex and fascinating
biological phenomena. It also  shows a significant potential to
deal with problems that combine complex fluid motion and molecule
dynamics. The efficiency of this scheme is also based on its
relatively low computational demand. The molecular part is largely
independent on the length of the molecule; the cost of the
mesoscopic (LB) part is known to scale linearly with the volume
occupied by the solvent. The {\it linear} scaling of the CPU time
with the molecular size (at constant volume) is the key feature of
the LB-MD approach, which permits the exploration of long
biomolecules ($N > 1000$) with a relatively modest computational
cost. Nevertheless, resort to parallel computing is mandatory, and
we expect the favorable properties of LB towards parallel
implementations to greatly facilitate this task. This will also
open the way to the simulation of systems at least an order of
magnitude larger than those considered so far, making it possible
to assign chemical specificity to the biopolymer constituents,
rather than the generic nature of the beads that constitute the
model polymers in the present study.  Work along these lines is in
progress.

\acknowledgments MF acknowledges support by Harvard's Nanoscale
Science and Engineering Center, funded by the National Science
Foundation, Award Number PHY-0117795.  SM and SS thank
Harvard's Initiative for Innovative Computing for its hospitality
and support.

%%%%%%%%%%%%%%%%%%%%%%%%%%%%%%%%%%%%%%%%%%


\begin{thebibliography}{99}

\bibitem{LBE}
D.~A. Wolf-Gladrow, ''Lattice gas cellular automata and lattice Boltzmann
models'', Springer Verlag, New York 2000;
S. Succi, ''Lattice Boltzmann Equation for Fluid Dynamics and Beyond'',
Oxford University Press, Oxford 2001;
R. Benzi, S. Succi, and M. Vergassola,
''The lattice Boltzmann-equation - Theory and applications'',
Phys. Rep., vol. 222, Dec. 1992, pp.~145--197.

\bibitem{LBEORIG}  G.~McNamara, G.~Zanetti,
''Use of the Boltzmann-equation to simulate lattice-gas automata'',
 Phys. Rev. Lett., vol. 61, no. 20, Nov. 1988, pp.~2332--2335;
 F.~Higuera, S.~Succi, and R.~Benzi,
''Lattice gas-dynamics with enhanced collisions",
 Europhys. Lett., vol. 9, no. 4, Jun. 1989, pp.~345--349;
 F.~Higuera, J.~Jimenez,
''Boltzmann approach to lattice gas simulations",
 Europhys. Lett., vol. 9, no. 7, Aug. 1989, pp.~663--668;
 H.~Chen, S.~Chen, W.~Matthaeus,
''Recovery of the Navier-Stokes equations using a lattice-gas Boltzmann
method", Phys Rev A, vol. 45, no. 8, Apr. 1992, pp.~R5339--R5342;
 Y.~H.~Qian, D.~d'Humieres, and P.~Lallemand,
''Lattice BGK models for Navier-Stokes equation",
 Europhys. Lett., vol. 17, no. 6, Feb. 1992, pp.~479--484;
 I.V.~Karlin, A.~Ferrante, H.C.~\"Ottinger,
''Perfect entropy functions of the Lattice Boltzmann method,
 Europhys. Lett., vol. 47, no. 2, Jul. 1999, pp.~182--188.

% \bibitem{LGA05} Proceedings of ''Discrete Simulation in Fluid Dynamics
%  2005'', edited by B.~Boghosian, special issue, Physica A, 362 (2006).

\bibitem{MULBE} S.~Succi, O.~Filippova, G.~Smith and E.~Kaxiras,
''Applying the lattice Boltzmann equation to multiscale fluid problems",
Comput. Sci. Eng., vol. 3, no. 6, Nov-Dec 2001, pp.~26--37.

\bibitem{ourLBM}
M.~G. Fyta, M. Melchionna, E. Kaxiras, and S. Succi,
''Multiscale coupling of molecular dynamics and hydrodynamics:
application to DNA translocation through a nanopore'',
Multiscale Model. Sim., vol. 5, no. 4, Dec. 2006, pp.~1156--1173.

\bibitem{DUN}  P.~Ahlrichs and B.~Duenweg,
''Lattice-Boltzmann simulation of polymer-solvent systems,
 Int. J. Mod. Phys. C, vol. 9, no. 8, Dec. 1998, pp.~1429--1438;
''Simulation of a single polymer chain in solution by combining
lattice Boltzmann and molecular dynamics'', J. Chem. Phys., vol. 111,
no. 17, Nov. 1999, pp.~8225--8239;
A.~Chatterji, J.~Horbach,
''Combining molecular dynamics with Lattice Boltzmann: A hybrid method
for the simulation of (charged) colloidal systems, J. Chem. Phys.,
vol. 122, no. 18, May 2005, Art. No. 184903.

\bibitem{LADDVERBERG} A.J.C.~Ladd and R.~Verberg. 
''Lattice-Boltzmann simulations of particle-fluid suspensions". 
J. Stat. Phys., vol. 104, (2001) pp.~1191-1251.

\bibitem{CATES} M.~E.~Cates, K.~Stratford, R.~Adhikari, P.~Stansell, J.~C.~Desplat, E.~Pagonabarraga 
and A.~J.~Wagner,
``Simulating colloid hydrodynamics with Lattice Boltzmann methods'',
J.Phys.Condensed Matter, vol. 16, Aug. 2004, pp.~S3903--S3910.

\bibitem{PREPA} S. Melchionna, J. Russo, S. Succi, 
''Lattice Boltzmann simulation of hydrodynamic few-body correlations'',
in preparation.


\bibitem{SHAKE} J.~P.~Ryckaert, G.~Ciccotti, and H.~J.~C.~Berendsen,
''Numerical-integration of cartesian equations of motion of a system with
constraints - Molecular-Dynamics of n-alkanes", J. Comp. Phys., vol. 23,
no. 3, 1977, pp.327--341.

\bibitem{RATTLE} H.~C.~Andersen,
''Rattle - A velocity version of the SHAKE algorithm for Molecular-Dynamics
calsulations", J. Comput. Phys., vol. 52, no. 1, (1983), pp.~24--34.


\bibitem{adhikari}R.~Adhikari, K.~Stratford, M.~E.Cates and A.J.~Wagner,
``Fluctuating Lattice Boltzmann'',
Europhys.Lett., vol. 71, no. 3, Aug. 2005, pp.~473--477.

\bibitem{karlin} S. Ansumali, I.V. Karlin and H.C. H\"ottinger,
''Minimal entropic kinetic models for hydrodynamics",
Europhys. Lett., vol. 63, no. 6, Sep. 2003, pp.~798-804;
 S. Ansumali and I.V. Karlin,
''Consistent Lattice Boltzmann method",
Phys. Rev. Lett., vol. 95, no. 16, Dec. 2005, Art. No. 260605.

\bibitem{melchJCP} S. Melchionna,
''Design of quasi-symplectic propagators for Langevin dynamics",
J. Chem. Phys., in press. 

\bibitem{EXPRM}
J.~J. Kasianowicz, E. Brandin, D. Branton, D., and D.~W.~L Deamer,
''Characterization of individual polynucleotide molecules
using a membrane channel'',
 Proc. Nat. Acad. Sci. USA, vol. 93, no. 24, Nov. 1996, pp.~13770--13773;
A. Meller, L. Nivon, E. Brandin, J. Golovchenko, and D. Branton,
''Rapid nanopore discrimination  between single polynucleotide molecules'',
vol. 97, no. 3, Feb. 2000, pp.~1079--1084;
J. Li, M. Gershow, D. Stein, E. Brandin, and J.~A. Golovshenko,
''DNA molecules and configurations in a solid-state nanopore microscope'',
Nat. Mater., vol. 2, no. 9, Sep. 2003, pp.~611--615.

\bibitem{TRANSL}
H. Lodish, D. Baltimore, A. Berk, S. Zipursky,
P. Matsudaira, and J. Darnell,
''Molecular Cell Biology'', W.H. Freeman and Company, New York, 1996.

\bibitem{statisTrans}
W. Sung and P.~J. Park,
''Polymer translocation through a pore in a membrane'',
 Phys. Rev. Lett., vol. 77, no. 7, Feb. 1996, pp.~783--786.

\bibitem{DynamPRL}
S. Matysiak, A. Montesi, M. Pasquali, A.~B. Kolomeisky, and
C. Clementi,
''Dynamics of polymer translocation through nanopores:
Theory meets experiment'', Phys. Rev. Lett., vol. 96, no. 11,
Mar. 2006, Art. No. 118103.

\bibitem{coarse}
D.~K. Lubensky and D.~R. Nelson,
''Driven polymer translocation through a narrow pore''
Biophys. J., vol. 77, no. 4, Oct. 1999, pp.~1824--1838 ;
Y.~Kantor and M.~Kardar,
''Anomalous dynamics of forced translocation'',
 Phys. Rev. E, vol. 69, no. 2, Feb. 2004, Art. No. 021806.

\bibitem{beadParam1} A.~J. Spakowitz and Z-G. Wang,
''DNA packaging in bacteriophage: Is twist important?",
Biophys. J., vol. 88, no. 6, Jun. 2005, pp.~3912--3923;
C. Forrey and M. Muthukumar,
''Langevin dynamics simulations of genome packing in bacteriophage",
Biophys. J., vol. 91, no. 1, pp. 25--41 (2006) and references therein.

\bibitem{beadParam2} T.~T.~Perkins, D.~E.~Smith, S.~Chu,
''Single polymer dynamics in an elongational flow'',
 Science, vol. 276, no. 5321, Jun. 1997, pp.~2016--2021;
 J.~S.~Hur, E.~S.~G.~Shaqfeh, and R.~G.~Larson,
''Brownian dynamics simulations of single DNA molecules in shear flow",
 J. Rheol., vol. 44, no. 4, Jul.-Aug. 2000, pp.~713--742;
 R.~M.~Jendrejack, J.~J.~de~Pablo, and M.~D.~Graham,
''Stochastic simulations of DNA in flow: Dynamics and the
effects of hydrodynamic interactions'',
J. Chem. Phys., vol. 116, no. 17, May 2002, pp.~7752--7759.

\bibitem{lpWCL}
Yamakawa, H. ''Modern Theory of Polymer Solutions'', Harper \& Row,
New York, 1971.

\bibitem{lpDNA}
P.~J. Hagerman,
''Flexibility of DNA'', Annu. Rev. Biophys. Biophys. Chem.,
vol. 17, 1988,  pp.~265--286;
S. Smith, L. Finzi, and C. Bustamante,
''Direct mechanical measurement of the elasticity of single DNA
molecules by using magnetic beads'',
Science, vol. 258, no. 5085, Nov. 1992, pp.~1122--1126.

\bibitem{NANO}
A.~J. Storm, C. Storm, J.~H. Chen, H. Zandbergen, J.~F. Joanny, and C. Dekker,
''Fast DNA translocation through a solid-state nanopore'',
Nanolett., vol. 5, no. 7, Jul. 2005, pp.~1193--1197.

\bibitem{blob}M. Fyta, S. Melchionna, S. Succi, and E. Kaxiras,
``Multiscale simulations of a translocating biopolymer through a
nanopore: Evidence for hydrodynamic correlations on molecular
motion'', under review.



\end{thebibliography}
\end{document}